# The break of shielding current at pulsed field magnetization of superconducting annuli.

V. S. Korotkov [1], E. P. Krasnoperov [1,2], A. A. Kartamyshev [1]

[1] National Research Center "Kurchatov Institute", 123182, Moscow
[2] Moscow Physical-Technical Institute, 141700, Dolgoprudniy
e-mail : vasmephi@mail.ru

**Abstract**

*The superconducting current breaking effect at pulsed field magnetization of HTS annuli was carefully investigated. It is proposed the simple model for qualitative description of this process. In particular, it is shown the appearance of a narrow sector of the considerable heating through which the magnetic flux penetrates into the superconducting annuli.*

**Introduction.**

Hard superconductors of type-II have the thermomagnetic instability at low temperatures. This instability manifested as avalanche-like flux penetration of magnetic field inside the superconductor [1]. It is so called flux jumps, appearing in the isothermal conditions, at any low perturbation. The thermomagnetic instability in high temperature superconductors (HTS) is not observed above 40 K due to a large heat capacity [2]. At high temperatures under the conditions of isothermal magnetization of a superconductor hollow cylinder or annulus (the ring with rectangular cross section) an inhomogeneous field is formed inside the body according to the critical state model, so $dH/dr \approx 4\pi J_C$ [1]. The superconducting current shields the interior until it fills the entire cross-section of the annulus. At the further magnetization the circulating current does not change its value maintaining the difference $(H_{ex}-H_{in})/(R_{ex}-R_{in}) = 4\pi J_c$ between the external and internal field.

In the case of pulsed field magnetization of superconducting annuli made of melt grown HTS the break of shielding current (BSC) followed by flux jump inside the annual's hole were observed at T = 78 K [3]. The dynamics of these flux jumps differs from classical thermomagnetic instability. The BSC appears only in case of high rate of external magnetic field increasing. In general the external field ramp rate dH/dt should be high enough that the generated power is larger than the heat flow into the surrounding liquid nitrogen. In this case, the BSC is initiated by the "weak spot" of the superconductor. Weak spots exist due to azimuthal anisotropy of critical current in the crystal (in a-b plane) [4], or due to the technological reasons during the HTS sample manufacturing. The increase of the resistance on this part of the annulus leads to the reduction of the dissipated power in the rest of the sample. As a result, the magnetic flux penetrates into the annuli through the narrow high resistive region. The size of this resistive region can be determined from the ratio of average and local temperature. For the annuli with external diameter of 30 mm the flux penetration region occupied the sector with an angle less than 10 degrees. The penetration time of magnetic flux was 0.3–0.5 ms [3].

In the present work the conditions of BSC were carefully investigated. For the qualitative description of this process we carried out the model calculation of pulse magnetization of the thin superconducting ring.

**Experimental technique and results**

For the experiments the melt grown annuli YBCO were used. The inner and outer diameters of the typical sample were 52 and 28 mm and thickness of 11 mm. Pulsed field magnetization technique and measurements of the circulated shielding current (with Rogowsky belt) were described in [3]. The experiments were performed under the liquid nitrogen temperature. After each pulse magnetization the superconductor was heated above $T_C$.

In fig. 1 the evolution of the shielding currents in the annulus for 3 pulses are presented. The duration of pulses was 35 ms, and the amplitudes $\mu_0 H_a$ were 0.92 T, 1.38 T, 2.7 T. The sharp drop in the current at $\tau_m$ is the break of shielding current. The observed picture of the BSC is similar as in [3]. The discrepancy in parameters of BSC are related to the size of the annulus (inductance L) and the value of the critical current. It is seen that with increasing $H_a$ (curves a-b-c) the BSC happens earlier ($\tau_m$ is reduced) and the maximal current value $I_m$ increases. The $I_m$ vs $H_a$ is shown in the inset of fig. 1. At the low amplitudes the field in the hole is shielded by the induced current. Therefore, in this range $I_m$ depends on $H_a$ linearly. Above a certain "critical" value of the magnetic field the BSC is observed and the magnetic flux abruptly penetrates into the annulus (flux jumps). In these conditions as it is shown in fig.1 the $I_m$ weakly increases with the amplitude of the external field. This behavior will be discussed below.

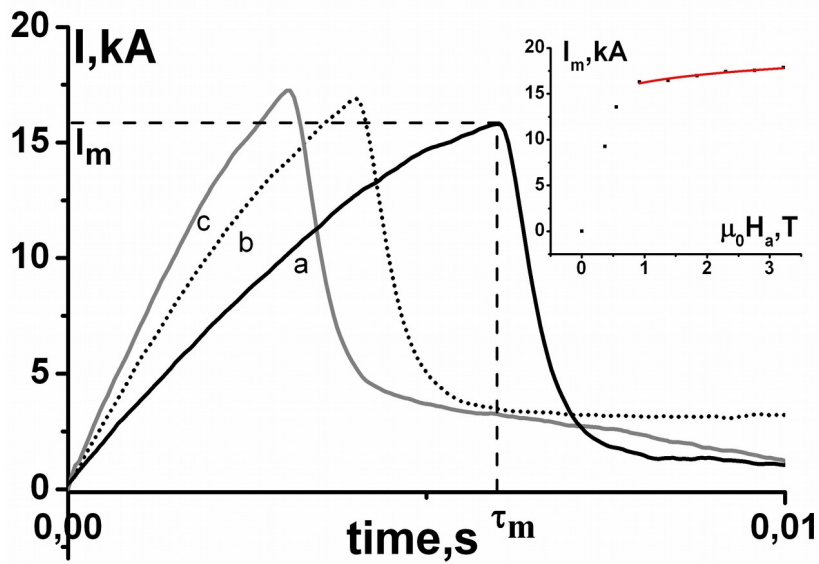

Fig. 1. The evolution of the shielding current for pulses durations of 35 ms with amplitudes $\mu_0 H_a$: 0.92 T (a), 1.38 T (b), 2.7 T (c). In the inset: (•) – the experimental data $I_m(\tau_m(H_a))$, solid curve – approximation function $(I_m)^{N+1} \propto 1/\tau_m$ for N=9.

Obviously, $I_m$ is indirectly connected to $H_a$. In fact, the BSC depends on the actual field value and local temperature, which, in turn, depends on the rate of the field's variation $dH/dt$ and $I_C$ distribution. In fig. 2 the evolutions of the shielding currents for two magnetizing pulses are shown. The magnetizing pulses have the same ratio of the amplitude $H_a$ to the pulse duration $\tau_p$, resulting in nearly the same $dH/dt$ for a long time. One can see that at the equal ramp rates, the shielding current evolution curves practically coincide near the BSC.

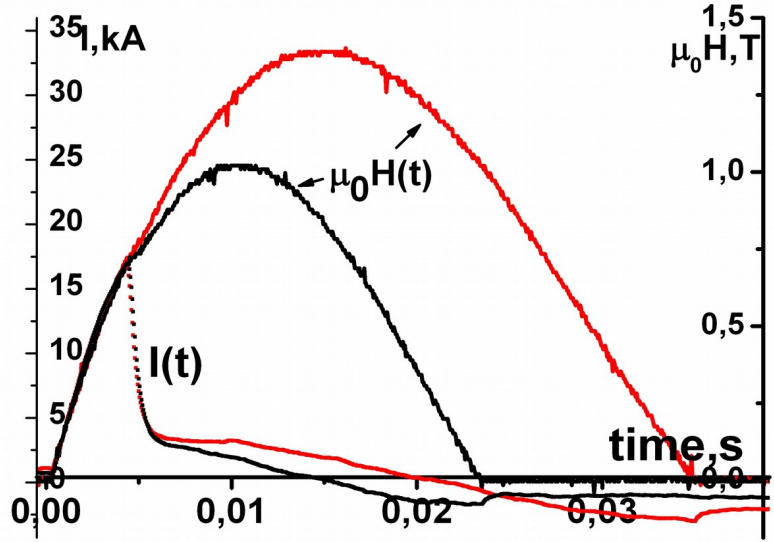

Fig. 2. Left axis – the evolution of the shielding current I(t) at the equal field ramp rate dH/dt. Right axis – oscillograms H(t) of the external magnetic field.

By analogy with [3] the I(t) dependence at BSC depicted in fig.2 allows estimating the size of overheated annulus region. Drop of current is approximated by exponential function with characteristic time $\tau_f = L/R_{flow}$. Here $L = 2.95 \cdot 10^{-8}$ H – annulus inductance calculated from [5], $R_{flow}$ – flux flow resistance. The annulus resistance evaluated from the characteristic time is $R_{flow} = 50$ μOhm.

The heat dissipated during BCS is $q_j = \int I^2 \cdot R_{flow}\, dt \approx 5$ J. If this energy was distributed on the whole volume of the annulus, the average temperature increase would not exceed $\langle \Delta T \rangle \approx 0.35$ K. 6–fold decrease of the shielding current at BSC corresponds to the local overheating $\Delta T_j \approx 9$ K. The volume of the annulus occupied by the moving fluxoid during BSC is $\Omega_{flow} = q_j / (C \cdot \Delta T_j)$, where C is the volumetric heat capacity [6]. In this case the width of the resistive hot region can be estimated as a sector with an angle of $\varphi = 2\pi \cdot \langle \Delta T \rangle / \Delta T_j \approx 15^0$.

**Calculation of the shielding current in the thin ring.**

For a qualitative description of the BSC process we considered the pulse magnetization of a thin superconducting ring. It was supposed that:
1. The ring has small cross-section, and the current is distributed homogeneously;
2. The diffusion time of the magnetic field is small;
3. The adiabatic approximation is valid, because the real characteristic time of cooling of the sample in liquid nitrogen $\tau_q = 1 - 2$ s [7] is two orders of magnitude higher than pulse duration ($\tau_p = 35$ ms) and 3 orders of magnitude higher than duration of flux jump.

The evolution of current in thin ring with an inductance L and resistance r is described by the well-known equation $L \cdot dI/dt = -I \cdot r + \varepsilon$. Here $\varepsilon = -S \cdot dH/dt$ is emf induced by magnetizing field. In the case of the superconductor the calculations are complicated by the fact that the resistance $r = U/I$ is a nonlinear function of the current which can be approximated by the power-law current voltage characteristic (CVC) is ordinarily used [8]. In the result we have:

$$L \cdot dI/dt = -U_0 (I/I_c)^N + \varepsilon \qquad (1)$$

Usually for the bulk HTS the exponent N is in range from 6 [9] and 21 [10]. In our experiment exponent N could be estimated from the dependence $I_m(H_a)$ presented in insert of fig.1. At the pulsed magnetization, the shielding current drops due to the heating which in turn depends on CVC. Formally, as one can see from (1), at $I = I_m$, $dI/dt = 0$. This results in $U_0(I/I_c)^N = \varepsilon$. The local high resistive state appears when the current in the "weak spot" exceeds the critical value. In this place the heat is predominantly dissipated. The conditions for BSC can be formulated as a dissipation of the heating energy $q = \int I \cdot U \cdot dt$ in the weak spot, which is enough to rise the temperature to $T_C$. The maximum current $I_m$, after which the BSC happens one can estimate from the condition $I_m \cdot U \approx q/\tau_m$. The relationship between the maximal current and the time to the current break $(I_m)^{N+1} \propto 1/\tau_m$ can be obtained using $U = U_0(I/I_c)^N$. This allows evaluating the N by fitting the dependence $I_m(\tau_m(H_a))$ shown in the inset of fig 2. For different annuli investigated in this work N was in the range of 8 – 10.

In our calculations we assumed N = 9, as defined above. For the primary calculations $U_0 = 10^{-7}$ V (E = $10^{-7}$ V/m) and $I_{C0}$ = 10 kA ($J_{C0}$ = 10 kA/cm$^2$) were set. These E and $J_{C0}$ are typical values for the electric field and current densities in melt grown HTS. On the final stage of the calculation, the critical current value $I_{C0}$ was adjusted to improve the agreement with the experiment.

The "weak spot" results in inhomogeneous azimuthal distribution of the critical current in the annuli $I_C = I_C(\varphi)$. We placed the reference point $\varphi = 0$ in the center of the weak spot and assumed $I_C(\varphi) = I_0 \cdot (1 - \alpha \cdot \exp(-\varphi^2/D^2))$. Here $\alpha$ is the "depth" of the inhomogeneity of current in the center of the weak spot, D – the "width" of the inhomogeneity (dispersion). Considering the linear decrease of the critical current with temperature one can write:

$$I_C(\varphi) = I_{C0}\left(1 - \alpha \cdot \exp\left(\frac{-\varphi^2}{D^2}\right)\right)\left(1 - \frac{T(\varphi) - T_0}{T_C - T_0}\right) \qquad (2)$$

$I_{C0}$ is the critical current at initial temperature $T_0$ = 77.5 K. Due to the inhomogeneity of $I_C$ the resistance depends on the angle $\varphi$. In this case eq. (1) should be written as

$$L\frac{dI(t)}{dt} = -\frac{U_0}{2\pi} \cdot I^N \cdot \int_{-\pi}^{\pi} \frac{1}{I_C^N(\varphi)} \cdot d\varphi + \varepsilon \qquad (3)$$

This equation is complicated by the fact that $I_C$ depends on the temperature, which is determined by the generated power related to the CVC of the superconductor and the shielding current value.

The evolution of the temperature in each element of the superconducting ring is determined by the Joule-Lenz law. The temperature increment is defined by the $dT = dt \cdot I \cdot U/C$, where C is the heat capacity. To solve this equation we used the implicit numerical method Cauchy-Euler [11].

**Procedure of the calculation.**

The ring was divided into M = 1000 identical elements. Every element $\varphi_i = i \cdot 2\pi/M$ is the sector of ring with angle interval $d\varphi = 2\pi/M$. We assume that within $d\varphi$ the parameters of the superconductor are unvarying. The current, temperature and voltage (from CVC) for elements $\varphi_i$ were calculated systematically for the each moment of time $t_j = j \cdot dt$. The time step dt

= 1 μs was chosen from the condition that it is small enough in comparison with the time of the BSC.

We introduce two-dimension arrays : temperature $\{T_{i,j}\} = T(\varphi_i,t_j)$, critical current $\{I_{Ci,j}\}= I_C(\varphi_i,t_j)$, and voltages $\{U_{i,j}\} = U(\varphi_i,t_j) = 1/M \cdot U_0 \cdot I_j^N/I_{Cj}^N(\varphi_i)$, which correspond to ring element $\varphi_i$ in the moment of time $t_j$, respectively. Also we introduce one-dimensional array $\{I_j\}$, which elements correspond to the current value in the ring at moments $t_j$.

To find the current in the moment $t_j$, the equation (3) were solved by the implicit Cauchy-Euler method. This method, as it turned, is more precise and stable for solving (3) than the explicit method [11]. In this method the difference (in this case, of the current) is written as half of sum of adjacent values of the function in the right part of the equation (3). The increment of the current was calculated in the assumption that during the time interval of $\Delta t = 1$ μs the temperature increment (and corresponding $I_c$) gives the contribution of the second order. So,

$$I_{j+1} - I_j = \frac{dt}{2} L \cdot \left( -U_0 \cdot I_{j+1}^N \cdot \sum_i \frac{1}{I_c^N(\varphi_i, t_j)} + E(t_{j+1}) - U_0 \cdot I_j^N \cdot \sum_i \frac{1}{I_c^N(\varphi_i, t_j)} + E(t_j) \right), \quad (4)$$

The equation (4) is an algebraic equation of the N-th degree of the current $I_{j+1}$. This one was solved numerically, by the Newton method. From the set of solutions it was selected only positive real root.

After the calculation of the current $I_{j+1}$ the dissipated energy in each element $\varphi_i$ and the temperature of each element were calculated from the relation $T_{i,j+1} = T_{i,j} + \Delta T_{i,j+1}$, where $\Delta T_{i,j+1} = U_{i,j+1} \cdot I_{j+1} \cdot \Delta t / C$. At $t = 0$ the current is absent ($I_0 = 0$). Using the value $T_{i,j+1}$ the critical current $I_{ci,j+1}$ was calculated with the factor of the anisotropy from Eq. (2). Using the value $I_{ci,j+1}$ the procedure of the calculation was repeated for the next current increment (Eq. (4)).

When the temperature approaches $T_C$ the critical current $I_C$ tends to zero, so the CVC ($U \propto (1/I_C)^N$) has a divergence near $T_C$. In this region the power law for the CVC turns into the Ohm's law. This fact complicates the algorithm of the calculation. The flux flow resistance was estimated in [3]. It was shown that the resistivity of a superconductor does not exceed the half of resistivity in normal state ($\rho_{flow} < \rho_n/2$). So we defined:

*if $r_{i,j} < r_n/2$, than $U_{ij} = U_0 (I_j^N/I_{cij}^N)$,*
*if $r_{i,j} > r_n/2$, than $U_{ij} = I_j r_n /2$.*

For the calculations of the BSC the following parameters were used :
- the azimuthal inhomogeneity is $D = 0.1$, $\alpha = 0.1$,
- the magnetization pulse amplitude is $\mu_0 H_a = 1.38$ T,
 T, duration $\tau_p = 35$ ms,
- the external field in the form of half-sine $\mu_0 H(t) = \mu_0 H_a \cdot \sin(\pi t/\tau_p)$,
- the normal resistance of the crystal YBCO at 92 K is $\rho_n = 100$ μOhm·cm,
- volumetric heat capacity $C = 0.75$ J/(K·cm$^3$) [6].

The calculated shielding current $I(t)$ was compared with the experimental curve and then the initial value of the critical current of CVC was corrected. The value $J_{C0}$ was changed to make a calculated $I(t)$ as close as possible to experimental curve. The result of calculation for the CVC with corrected $I_{C0} = 8.5$ kA is shown by the dashed curve in the fig 3. On the initial part of the curve the critical current is high and the induced current completely screens the external field. With further increasing of the shielding current the right part of the equation (1) turns into zero and the shielding current approaches the maximal value.

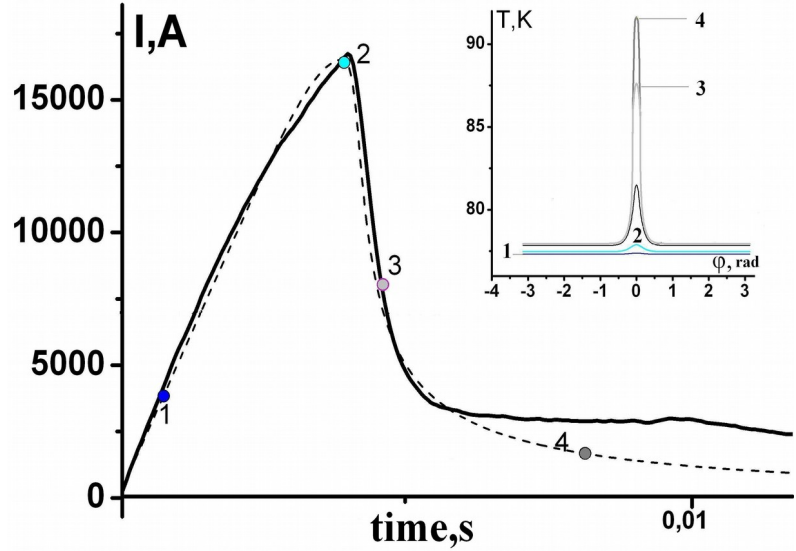

Fig 3. The evolution of the induced current I(t) at the pulse $\mu_0 H_a = 1.38$ T: in the experiment (solid curve) and calculated (dashed curve). In the inset : the azimuthal distribution of temperature at the moments of time marked by corresponding numbers on the I(t) curve.

Since the current $I_j$ is common for all elements of the ring, the generated heat (and temperature increment) is maximal in i-th elements where the critical current value $I_{ci,j}$ is minimal. The temperature in these elements rise and it leads to further decrease of the $I_C$ value. As a consequence, the heat dissipation increases there. Thereby, the rise of the local temperature and the corresponding drop of $I_c$ develops in the ring's sector with reduced value of the critical current.

In the inset of fig. 3 the azimuthal profiles of the temperature $T(\varphi)$ at the different moments of time are shown. The numbers 1–4 on the curve I(t) mark the time for which corresponding $T(\varphi)$ profiles (1–4) were calculated. It is seen that before the BSC (curves 1 and 2), the temperature profile is relatively flat. During the BSC (curves 3 and 4) the temperature precipitously rises up to the $T_c$ what accompanied by the current decrease by the order of magnitude.

After the BSC the distribution $T(\varphi)$ has a narrow peak in the vicinity of the inhomogeneity and the flat part far away from the inhomogeneity. The temperature in the maximum is higher than liquid nitrogen temperature by 12 K. The high resistive region (the character overheat width of the channel) is about $12^0$. In this experiment, as mention above, the temperature jump is $\Delta T_j \approx 9$ K, and width of the channel is $\delta\varphi \approx 15^0$.

We analyzed the influence of the parameters $\alpha$ and D from (2) on the temperature peak. It turned out that the increase of $\alpha$ from 0.01 to 0.2 practically does not affect the temperature distribution in the vicinity of the peak, but it lowers the flat part (baseline $T(\varphi)$) by $\approx 2.5$ K. At these conditions the BSC appears earlier ($\tau_m$ decreases). This result is obvious because in the case of small inhomogeneity of critical current (small $\alpha$) the local overheating develops slower and the average heating is higher. Calculations showed, as well, that the size of the inhomogeneity D does not affect the width of the channel $\delta\varphi$. So, with increasing D by 3 orders (from 0.01 to 10) the width of the channel increases by the factor of 2 (from $10^0$ to $23^0$).

This model demonstrates that the increasing of the pulse amplitude leads to a slight increase of the shielding current maximum and reduction of $\tau_m$. It is in a good agreement with the experimental results shown in fig. 1.

## Conclusion

During pulsed magnetization of multiply–connected superconductors such as ring or annulus the local heating in the region with lowest $J_c$ creates a narrow channel of the high resistive state. This leads to the break of the shielding current and to the jump of the magnetic flux into hole of the superconductor. Our model calculations of a pulse magnetization of thin ring are in a good agreement with experimental results on the superconducting annulus.

## Acknowledgments.

Authors are very grateful to V.Grinenko and V.Viatkin for criticism and proofreading of the draft text.